\documentstyle[epsfig]{qcdparis}

\def\0{\over } \def\1{\vec }     \def\2{{1\over2}} \def\4{{1\over4}}
\def\5{\bar }  \def\6{\partial } \def\7#1{{#1}\llap{/}}
\def\8#1{{\textstyle{#1}}}       \def\9#1{{\bf {#1}}}
 \def\llp{\hbox to 0pt{\hss /\hskip1.5pt}}
\def\llo{\hbox to 0.2pt{\hss /}} \def\llq{\hbox to 0pt{\hss /\hskip0.5pt}}
\def\so{\supset\hbox to 0pt{\hss $\displaystyle -$\hskip1pt}}

\def\<{\langle } \def\>{\rangle }

\def\bea{\begin{eqnarray}} \def\eea{\end{eqnarray}}
\def\beann{\begin{eqnarray*}} \def\eeann{\end{eqnarray*}}
\def\beq{\begin{equation}} \def\eeq{\end{equation}}

\begin{document}
\pagestyle{plain}
\title{Calculating the diffractive from the inclusive structure function}

\author{W. Buchm\"{u}ller and A. Hebecker}

\affil{Deutsches Elektronen-Synchrotron DESY, 22603 Hamburg, Germany}

\abstract{It is demonstrated that the global properties of the rapidity gap
events at HERA can be understood based on electron-gluon scattering and a
non-perturbative mechanism of colour neutralization. Using the measured
inclusive structure function $F_2$ to determine the parameters of the parton
model, the diffractive structure function $F_2^D$ is predicted. The ratio of
diffractive and inclusive cross sections, $R_D = \sigma_D/\sigma_{incl}\simeq
1/9$, is determined by the probability of the produced quark-antiquark pair to
evolve into a colour singlet state.}

\resume{Les \'ev\'enements \`a gap de rapidit\'e observ\'es \`a HERA peuvent
\^etre attribu\'es \`a la diffusion d'un gluon et d'un \'electron combin\'e
avec un m\'echanisme non-perturbative pour la neutralisation de la couleur.
Nous utilisons la fonction de structure inclusive $F_2$ pour d\'eterminer les
param\`etres du mod\`ele de partons et pour pr\'evoir la fonction de structure
diffractive $F_2^D$. Le quotient des sections efficaces diffractives et
inclusives, $R_D=\sigma_D/\sigma_{incl}\simeq 1/9$ correspond \`a la
probabilit\'e que le pair quark-antiquark produit dans ce proc\`es \'evolue
dans un \'etat neutre par rapport \`a la couleur.}

\twocolumn[\maketitle]
\fnm{7}{Talk given in the Session on Diffraction at the Workshop on Deep
Inelastic Scattering and QCD, Paris, April 1995}
\noindent Recently, detailed analyses of rapidity gap events in deep
inelastic scattering at HERA have appeared confirming the previously observed
leading twist behaviour of the diffractive cross section
\cite{h1}-\cite{zeus}. The absence of a hadronic energy flow between proton
remnant and current fragment suggests that in the scattering process a colour
neutral part of the proton is stripped off. A leading twist behaviour is
usually regarded as evidence for scattering on point-like constituents. This,
however, appears to be in conflict with the fact that quarks and gluons carry
colour.

In the following we shall demonstrate that this puzzle can be resolved
considering the production of a quark-antiquark pair in electron-gluon
scattering as basic partonic process and taking non-perturbative
fragmentation effects into account \cite{bu3}. Our approach is related to
previous work on ``aligned jet models'' \cite{strikman} and ``wee parton
lumps'' in deep inelastic scattering \cite{bu1,bu2}. Photon-gluon fusion has
been independently considered as basic process underlying rapidity gap events
by Kowalski \cite{kowalski} and Edin et al. \cite{edin}.

The relevant kinematic variables are defined in \fref{fig1}.
\ffig{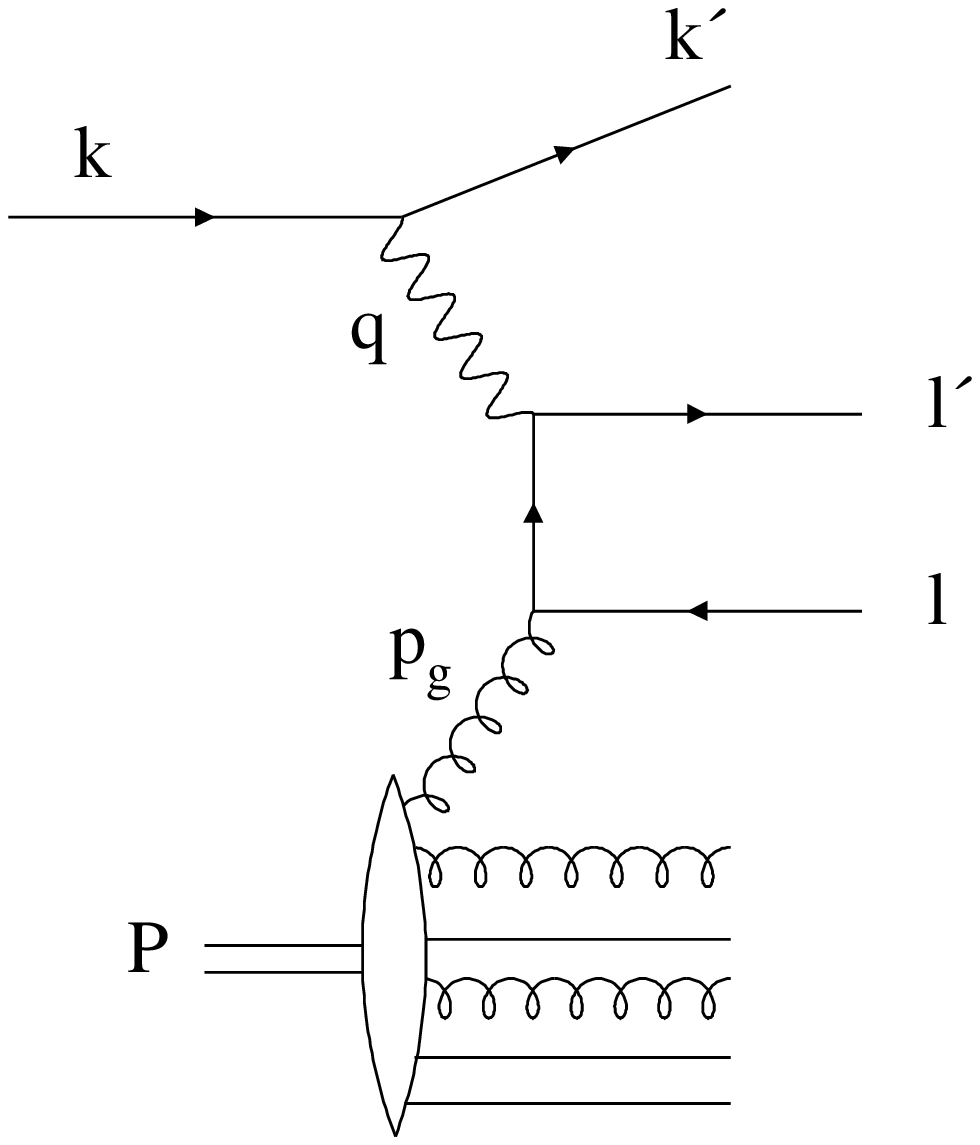}{50mm}{\em Quark-antiquark pair production in electron-gluon
scattering.}{fig1}
With $\vec{p}_g = \xi \vec{P}$ , $-p_g^2 = m_g^2 \ll Q^2$ and the invariant
mass square $M^2=(q+p_g)^2$ one has,
\beq
\beta \equiv {Q^2\over Q^2+M^2} \simeq {x\over \xi}\ .
\eeq

The differential cross section for the inclusive production of quark-antiquark
pairs is given by \cite{field}
\beq\label{xep}\hspace*{-5cm}
{d\sigma(ep\rightarrow e(q\bar{q})X)\over dx dQ^2 d\xi}=
\eeq
\[
= {\alpha\over \pi x Q^2} g(\xi)\left(\left(1-y+{y^2\over 2}\right)
\left(\sigma_T+\sigma_L\right)-{y^2\over 2}\sigma_L\right)\ .
\]
Here $g(\xi)$ is the gluon density, and the cross section $\sigma_{T(L)}$ is
obtained by integrating the differential parton cross section over the
kinematically allowed range for the momentum transfer $t=(q-l')^2$. The
differential partonic cross sections for the absorption of a virtual photon
$\sigma_T$ and $\sigma_L$, calculated in the massive gluon scheme, can be
found in ref.~\cite{field}.

At small values of $\xi$ the quark and antiquark densities are much smaller
than the gluon density. Therefore we shall neglect the quark contribution to
$F_2$ in the following. Note, that in our model this essentially amounts to
calculating the sea quark densities at scale $Q$ in terms of a gluon density
$g(\xi)$ at a scale $m_g = {\cal O}$(1 GeV).

{}From eq.~(\ref{xep}) together with the virtual photon cross sections one
then obtains the inclusive structure function $F_2(x,Q^2)$,
\beq\label{f2g}\hspace*{-2cm}
F_2(x,Q^2)= x{\alpha_s\over 2\pi} \sum_q e_q^2
\int_x^1{d\xi\over \xi}g(\xi)\times
\eeq
\[
\times\left((\beta^2+(1-\beta)^2)\ln{Q^2\over m_g^2 \beta^2}
- 2 + 6\beta(1-\beta) \right)\ .
\]
The virtuality $-m_g^2$ of the gluon regularizes the collinear divergence at
$t=0$.

For the gluon density at small values of $\xi$ we use the usual
parameterization,
\beq\label{gluon}
g(\xi)\ =\ A_g\ \xi^{-1-\lambda}\ ,
\eeq
where $A_g$ is a constant. Inserting eq.~(\ref{gluon}) into eq.~(\ref{f2g})
we can now evaluate the inclusive structure function $F_2$. In the small--$x$
region the obtained expression can be further simplified for small values of
the exponent $\lambda$. This yields
\beq\label{fit}
F_2(x,Q^2) \simeq {\alpha_s\over 3\pi}\sum_q e_q^2 x g(x)
\left({2\over 3} + \ln{Q^2\over m_g^2}\right)\ .
\eeq
For the parameters $\lambda=0.23$, $m_g=1.0$ GeV and $A_g \alpha_s \sum_q
e_q^2 = 0.61$ the above expression provides a good description of the H1
measurement of the structure function (compare the phenomenological fit of
$F_2$ in \cite{h1f2}).

We are now ready to calculate the diffractive structure function. The main
idea is that the quark-antiquark pair, originally produced in a colour octet
state, changes its colour through further soft interactions with the
colour field of the proton remnant. Hence, the quark-antiquark pair evolves
into a parton cluster which separates from the proton remnant with some
probability $P_8$ in a colour octet state, and with probability $P_1\!=\!1\!-
\!P_8$ in a colour singlet state. In the first case a colour flow
between proton remnant and current fragment leads to the typical hadronic
final state. In the latter case, however, the colour singlet final state
fragments independently of the proton remnant, yielding a gap in rapidity. For
a sufficiently fast rotation of the colour spin of quark and antiquark, the
probabilities should simply be given by the statistical weight factor
accounting for the possible states of the quark-antiquark pair, i.e.,
$P_1 \simeq 1/9\ , P_8\simeq 8/9$.

Similar ideas concerning the rotation of quarks in colour space have
been discussed by Nachtmann and Reiter \cite{nacht} in connection with
QCD-vacuum effects on hadron-hadron scattering. Another approach describing
the colour rotation by a soft gluonic field is based on the eikonal
approximation to the quark propagator \cite{nacht1}. Here a non-abelian phase
is introduced through a path ordered integral along the fermion line.

The diffractive structure functions are defined as
\beq
{d\sigma_D \over dx dQ^2 d\xi} = {4\pi \alpha^2 \over x Q^4}\!
\left(\!\left(1 - y + {y^2\over 2}\right)F_2^D-{y^2\over 2} F_L^D\right).
\eeq
$F_2^D(x,Q^2,\xi)$ is easily obtained from eq.~(\ref{f2g}). With $x=\beta\xi$,
and including the statistical weight factor $P_1$, one obtains
\beq\label{f2d}
F_2^D(x,Q^2,\xi)\simeq{1\over 9}{\alpha_s\over 2\pi} \sum_q e_q^2
g(\xi)\bar{F}_2^D(\beta,Q^2)\ ,
\eeq
where\vspace{-.2cm}

\[
\bar{F}_2^D(\beta,Q^2)\!=\!\beta\!\left(\!(\beta^2+\!(1\!-\!\beta)^2)\ln{Q^2
\over m_g^2 \beta^2}\!-\!2\!+ 6\beta(1\!-\!\beta)\!\right).
\]
Since the gluon density $g(\xi)$ and the mass scale $m_g$ have been determined
by the fit to the inclusive structure function $F_2$, the diffractive
structure function is unambiguously predicted, including its normalization.

As an immediate consequence the ratio of diffractive and
inclusive cross sections follows from eqs.~(\ref{f2g}) and (\ref{f2d}),
\beq
R_D ={\int_x^1 d\xi F_2^D(x,Q^2,\xi)\over F_2(x,Q^2)} \simeq {1\over 9}\ .
\eeq
This ratio directly measures the probability of forming a colour singlet
parton cluster in the scattering process.

The form of the diffractive structure function (\ref{f2d}) is identical with
expressions obtained based on the idea of a ``pomeron structure function''
\cite{ing,pom}. The interpretation of the ingredients, however, is rather
different. The ``pomeron flux factor'' is replaced by the density of gluons
inside the proton, which factorizes. The ``pomeron structure function'' for
partons with momentum fraction $\beta$ inside the ``pomeron'' is identified
as the differential distribution for the production of a quark-antiquark pair
with invariant mass $M^2$.

As it can be seen from \fref{fig2}, the function $\bar{F}_2^D(\beta,Q^2)$ is
rather flat for intermediate values of $\beta$ between
$0.2$ and $0.6$.
\ffig{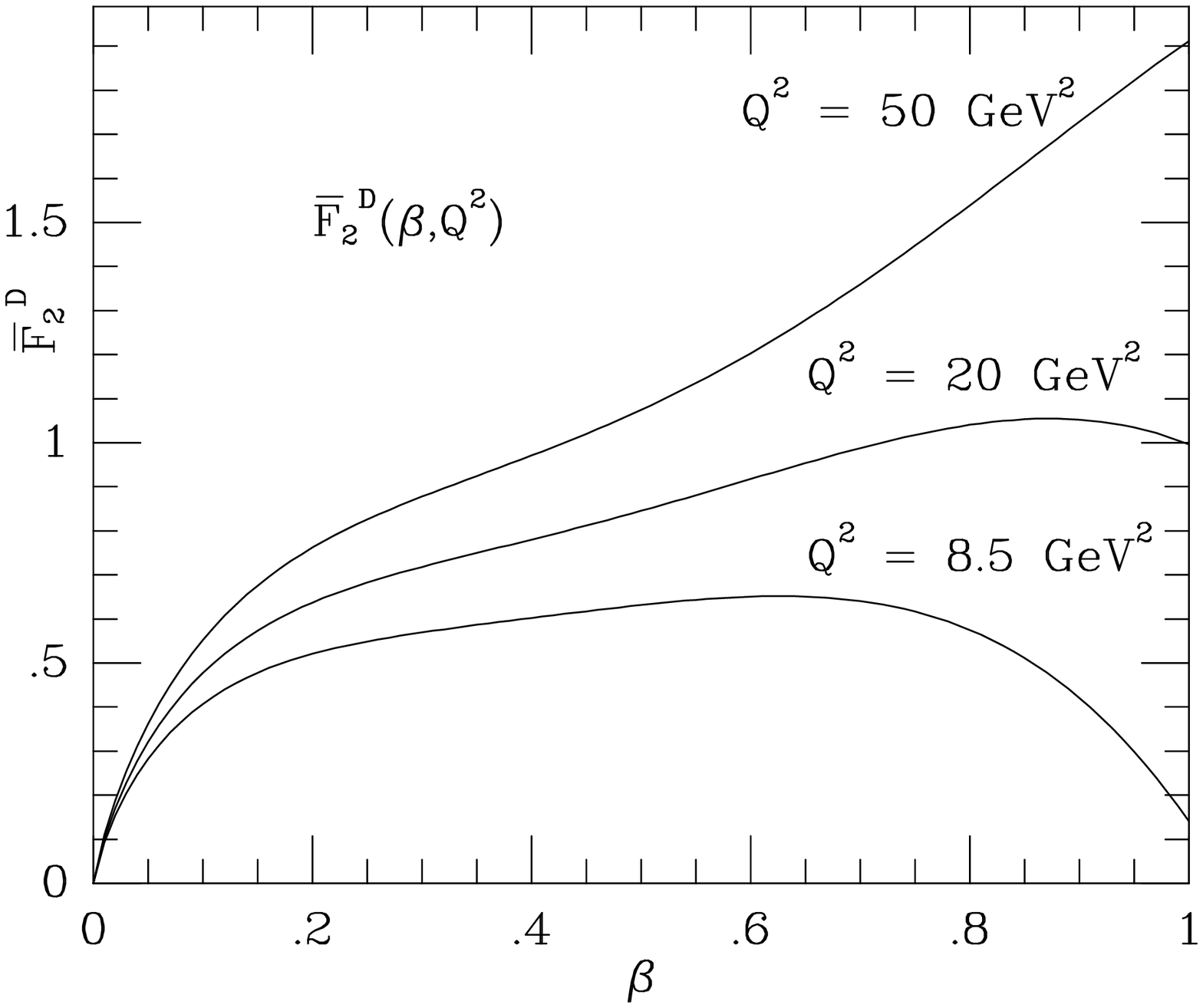}{66mm}{\em Dependence of the diffractive structure function on
$\beta$ and $Q^2$.}{fig2}
Approximating $\bar{F}_2^D(\beta,Q^2)$ in this interval by
$\bar{F}_2^D(0.4,Q^2)$, a comparison of eqs.~(\ref{f2g}) and (\ref{f2d})
yields the scaling relation \cite{bu2}
\beq\label{scaling}
F_2^D(x,Q^2,\xi) \simeq {0.04\over \xi} F_2(\xi,Q^2)\ .
\eeq
This scaling relation provides a rather accurate description \cite{bu3} of
recent measurements of the diffractive structure function by the H1
collaboration \cite{dainton}.

Let us finally verify the appearance of rapidity gaps within the above
model. The rapidity of the antiquark with momentum $l$ in the
$\gamma^* p$-rest frame is related to other kinematical
variables by
\beq\label{eta}
\eta = {1\over 2}\ln\left[\xi(1-\beta)\frac{u+m_g^2\beta}{t+m_g^2\beta}\right]
\, .
\eeq
Using this relation the differential cross section as a function of the
rapidity of the antiquark can be calculated from eqs.~(\ref{xep}) and
(\ref{eta}). The total diffractive cross section for a maximum rapidity
$\eta_{max}$ can now be obtained by integrating over the kinematic domain
where the rapidity of the antiquark is larger than the rapidity of the quark.
The reverse configurations yield the same contribution. Using the kinematic
boundaries of ref.~\cite{h1} the $\eta_{max}$-distribution has been calculated
in the described manner \cite{bu3}. Above $\eta_{max}\sim 2$ the diffractive
cross section is found to be negligible.

In summary, we have demonstrated that electron-gluon scattering can
account for the global properties of the rapidity gap events observed
at HERA provided the following two hypotheses are correct: First, the
initially produced quark-antiquark pair evolves with a probability $P_1\simeq
1/9$ into a colour singlet parton cluster; second, the rapidity range of the
diffractive hadronic final state is essentially given by the rapidity interval
spanned by the produced quark-antiquark pair.

This simple picture appears to provide a rather accurate description of
the observed diffractive events, including the total rate, the
$\xi$-dependence and the $Q^2$-dependence. However, a number of theoretical
issues still remain to be settled. Our results indicate, that a semiclassical
approach to the small-$x$ region might be appropriate, where ``wee partons''
are treated as a classical colour field.

\vspace{2cm}


\Bibliography{100}
%
\bibitem{h1}
H1 collaboration, T.~Ahmed at al., Phys.~Lett.~B348 (1995) 681
\bibitem{dainton}
J.~Dainton, these proceedings
\bibitem{foster}
B.~Foster, these proceedings
\bibitem{zeus}
ZEUS collaboration, M.~Derrick et al., DESY 95-093 (1995)
\bibitem{bu3}
W.~Buchm\"uller and A.~Hebecker, DESY 95-077, to appear in Phys. Lett. B
\bibitem{strikman}
M.~Strikman, these proceedings
\bibitem{bu1}
W.~Buchm\"uller, Phys.~Lett.~B335 (1994) 479
\bibitem{bu2}
W.~Buchm\"uller, Phys.~Lett.~B353 (1995) 335
\bibitem{kowalski}
H.~Kowalski, talk at the {\it Int.~Workshop on Deep Inelastic Scattering and
Related Subjects} (Eilat, 1994), unpublished
\bibitem{edin}
A.~Edin, G.~Ingelman and J.~Rathsman, these proceedings
\bibitem{field}
R.D.~Field, {\it Applications of Perturbative QCD}, Addison Wesley,
New York, 1989
\bibitem{h1f2}
H1 collaboration, T.~Ahmed et al., Nucl.~Phys.~B439 (1995) 471
\bibitem{nacht}
O.~Nachtmann and A.~Reiter, Z.~Phys.~C24 (1984) 283
\bibitem{nacht1}
O. Nachtmann, Ann.~Phys.~209 (1991) 436
\bibitem{ing}
G.~Ingelman and P.~Schlein, Phys.~Lett.~B152 (1985) 256
\bibitem{pom}
K.~Golec-Biernat; A.~Kaidalov; H.-G.~Kohrs;\\ P.V.~Landshoff; these
proceedings
\end{thebibliography}
\end{document}